\documentclass[12pt,a4paper,oneside,english,leqno]{article}
\usepackage{amsfonts,amsmath,amssymb}
\usepackage[mathscr]{euscript}
\usepackage[numbers]{natbib}
\usepackage{hyphenat}
\usepackage[numbib,nottoc]{tocbibind}

\usepackage[english]{babel}
\usepackage[latin1]{inputenc}
\usepackage[T1]{fontenc}
\usepackage{lmodern}
																
\newcommand{\us}[1][t]{ \ensuremath{U^{\uparrow}_{#1}}} 	%Upper Snell envelope
\newcommand{\ftp}{ \ensuremath{\mt}} 			% Family of stopping times
\newcommand{\tp}{ \ensuremath{\theta}}	 	% A tipical stopping time

\newcommand{\esssup}{ \ensuremath{\mathrm{ess \thinspace sup}}} 	%essential supremum

\newcommand{\fnd}[3][Definition]{
\newtheorem{d1}{#1}[section] 
\begin{d1}  \label{#2}
#3
\end{d1}
}

\newcommand{\nd}[3][Definition]{
\newtheorem{#2}[d1]{#1}
\begin{#2}  \label{#2}
#3
\end{#2}
}

\newcommand{\eq}[2]{\begin{equation} \label{#1} 
#2 \end{equation}}

\newcommand{\nl}[3][Lemma]{
\newtheorem{#2}[d1]{#1}
\begin{#2}  \label{#2}
#3
\end{#2}
}

\newcommand{\np}[3][Proposition]{
\newtheorem{#2}[d1]{#1}
\begin{#2}  \label{#2}
#3
\end{#2}
}

\newcommand{\nt}[3][Theorem]{
\newtheorem{#2}[d1]{#1}
\begin{#2}  \label{#2}
#3
\end{#2}
}

\newcommand{\nc}[3][Corollary]{
\newtheorem{#2}[d1]{#1}
\begin{#2}  \label{#2}
#3
\end{#2}
}

\newcommand{\nrem}[3][Remark]{
\newtheorem{#2}[d1]{#1}
\begin{#2}  \label{#2}
#3
\end{#2}
}

\newcommand{\fincor}{\ensuremath{\square}}  %fin corollary
\newcommand{\finlemma}{\ensuremath{\square}}			%fin lema	
\newcommand{\finpro}{\ensuremath{\square}}				%fin proposiciones
\newcommand{\fintheo}{\ensuremath{\square}}	%fin teoremas

\newcommand{\mbf}{\ensuremath{\mathbb{F}}}

\newcommand{\mbr}{\ensuremath{\mathbb{R}}}

\newcommand{\ma}{\ensuremath{\mathcal{A}}}

\newcommand{\mf}{\ensuremath{\mathcal{F}}}
\newcommand{\mg}{\ensuremath{\mathcal{G}}}

\newcommand{\mq}{\ensuremath{\mathcal{Q}}}
\newcommand{\mP}{\ensuremath{\mathcal{P}}}

\newcommand{\mt}{\ensuremath{\mathcal{T}}}

\normalsize

\begin{document}

%-Zusammenfassung / Abstract*-------------------------------------
%*********************************************************************************************************************************************
\title{Monitoring dates of maximal risk}
\maketitle
\begin{center}
\author{{\bf Trevi\~no-Aguilar Erick\footnote{Centro de Investigaci\'on en Matem\'aticas A.C., Guanajuato M\'exico. email trevino@cimat.mx}}}
\end{center}
\abstract{
Monitoring means to observe a system for any changes which may occur over time, using a monitor or measuring device of some sort. In this paper we formulate a problem of monitoring dates of maximal risk of a financial position. Thus, the ``systems'' we are going to observe arise from situations in finance. The ``measuring device'' we are going to use is a time-consistent measure of risk.\\
In the first part of the paper we discuss the numerical representation of conditional convex risk measures which are defined in a space $L^p(\mf,R)$ and take values in $L^1(\mg,R)$, for $p \geq 1$. This will allow us to consider time-consistent convex risk measures in $L^1(R)$.\\
In the second part of the paper we use a time-consistent convex risk measure in order to define an abstract problem of monitoring stopping times of maximal risk. The penalty function involved in the robust representation changes qualitatively the time when maximal risk is for the first time identified. A phenomenon which we discuss from the point of view of robust statistics.\\
}\\
\paragraph{Keyword:} Convex Risk Measures, Monitoring, Stopping times, Time-consistency, Upper Snell envelope.

%************************************************************************************************************************
\section[Introduction]{Introduction}\label{labelintroduction}
%************************************************************************************************************************
The word  ``monitoring'' produces more than 170 millions of results on internet. We believe this is a prompt for the relevance of the concept, but also for the variety of contexts and specific meanings where it appears. Monitoring means ``to observe a system for any changes which may occur over time, using a monitor or measuring device of some sort''. In this paper we formulate a problem of monitoring dates of maximal risk of a financial position. Thus, the ``systems'' we are going to observe, arise from situations in finance. The ``measuring device'' we are going to use is a time-consistent measure of risk.\\
Consider the following concrete financial situation. After making a loan, a bank decides whether to continue or to reduce risk. The bank may either sell out the loan or buy  insurance through a credit derivative. With a credit derivative, the  bank retains the loans control rights but no longer has an incentive to monitor; with loan sales, control rights pass to the buyer of the loan, who can then monitor, although in a less-informed manner. The trade-off between selling out a loan or using derivatives to hedge default risk is discussed by several authors; see e.g., Duffee and Zhou\cite{Duffeezhou}. The role of information in credit issuing has been discussed by Stiglitz and Weiss\cite{Stiglitzweiss}. The effect of monitoring in financial systems has been discussed by e.g., Mundaca\cite{Mundaca}. In this paper we assume that the decision of monitoring has been taken and a monitoring policy has been defined. \textit{We focus on how to determine the best time to act}. Our main goal is to show that such financial situations can be analyzed in the framework of dynamical convex risk measures.\\     

The paper consists of two parts. In the first part we present our measuring device: A time-consistent dynamical convex risk measure $\Phi$ in $L^1(R)$. In the second part we formulate, and solve, an abstract problem of monitoring dates of maximal risk.\\

The paper is organized as follows. The first part of the paper corresponds to Section \ref{labelsectiondynamicalcrm}. We discuss the numerical representation of conditional convex risk measures which are defined in a space $L^p(\mf,R)$, for $p \geq 1$, and take values in $L^1(\mg,R)$ (in this sense, \textit{real-valued}). In the literature it has been discussed the numerical representation of (static) convex risk measures beyond essentially bounded financial positions; see e.g., Biagini and Fritelli\cite{Biaginifritelli}, Cheridito and Li\cite{Cheriditoli}, Filipovi\'c and Svindland\cite{Filipovicsvindland}, Kaina and R\"uschendorf\cite{Kainaruschendorf}, Kr\"atschmer\cite{Kratschmer}, Ruszczy\'nski and Shapiro\cite{Ruszczynskishapiro}. In other direction, it has been discussed the assessment of risks taking explicitly new information into account, i.e., conditional convex risk measures; see e.g., Bion-Nadal\cite{Bionnadalconditional}, Theorem 3, Cheridito et al\cite{Cheriditodelbaenkupperdmrmbdtp}, Theorem 3.16, and,  Detlefsen and Scandolo\cite{Detlefsenscandolo}, Theorem 1.\\
In Subsection \ref{labelsectionrvccrm}, we discuss conditional convex risk measures beyond $L^{\infty}(\mf,R)$. The main result of this section is the robust representation Theorem \ref{labelcrobustrepresentationlocallyequivalent}. The first step to prove Theorem \ref{labelcrobustrepresentationlocallyequivalent} is the  Proposition \ref{labeltheoremrobustrepresntation}. We show that any lower semicontinuous, non necessarily real-valued,  conditional risk measure defined in $L^{p}(\mf,R)$ is representable. This is a well-known property in the space of essentially bounded functions $L^{\infty}(\mf,R)$, we present an extension to the space $L^{p}(\mf,R)$. An interesting aspect of  Proposition \ref{labeltheoremrobustrepresntation}, we believe, is that we follow  a different proof-strategy. In the literature, the construction of numerical representations of conditional risk measures is reduced to an application of the theory of numerical representation of (static) risk measures in $L^{\infty}(\mf,R)$. Here we use the Hahn-Banach hyperplane separating Theorem following the ideas of the original papers. We then conclude the proof of Theorem  \ref{labelcrobustrepresentationlocallyequivalent} with an exhaustion argument started by Halmos and Savage. It is true that these two techniques have been applied in the theory of robust representation of risk measures before, as we specify with more detail below. However, we give a substantially different presentation to extend known results. As by-product of this approach, in Theorem \ref{labelcrobustrepresentationlocallyequivalent}, we obtain  a robust representation which involves exclusively \textit{essentially bounded penalizations}.\\
In subsection \ref{labelsectionrealvaluedccrm}, we specialize to real-valued conditional risk measures. We show that real-valued conditional risk measures are continuous. To this end, we use ideas due to Biagini and Fritelli \cite{Biaginifritelli}, Theorem 2 (the extended Namioka-Klee Theorem). Thus, real-valued conditional risk measures are representable; see Theorem \ref{labeltheoremrepresentabilityofrealvaluedccrm}. Then we apply ideas due to Cheridito and Li\cite{Cheriditoli} and Kaina and R\"uschendorf\cite{Kainaruschendorf}, in order to show that any penalty function representing a real-valued conditional risk measure must be \textit{coercive}; see Theorem \ref{labeltheoremcoercivitypropertynecessity}.\\
In Subsection \ref{labelsectionrvccrmtimeconsistent}, we prove that the minimal representation of real-valued conditional risk measures keeps invariant if the risk measure is restricted from $L^p(\mf,R)$ to $L^{\infty}(\mf,R)$. This invariance property will allow us to consider time-consistent risk measures  in $L^1(R)$, which is going to be the ``measuring device'' in Section \ref{labelsectionexistenceofopst}.\\ 

The second part of the paper corresponds to Section  \ref{labelsectionexistenceofopst}. We use a time-consistent convex risk measure in $L^1(R)$ in order to define an abstract problem of monitoring stopping times of maximal risk. If risk is quantified by a time-consistent convex risk measure $\Phi=\{\rho_t\}_{t=0,1,\ldots,T}$, the maximal risk of a financial position with discounted payoff $H:=\{H_t\}_{t=0,1,\ldots,T}$ takes the form
\[
\sup_{\theta}\rho_0(H_{\theta}),
\]
the supremum is taken over the family of stopping times of the period of time $\{0,1,\ldots,T\}$. Thus, we may say that $\theta^{*}$ is a stopping time of maximal risk for the payoff $H$ if
\[
\rho_0(H_{\theta^*})=\sup_{\theta}\rho_0(H_{\theta}).
\]
We are going to show that \textit{time-consistency} is a sufficient condition for the existence of  stopping times of maximal risk; see Theorem \ref{rcace}. The (``convex'') upper Snell envelope \eqref{labelequationuppersnellenvelope} will play a key role. This concept is formulated by F\"ollmer and Schied\cite{Follmerschied} in the context of arbitrage free prices for  American options; see \cite{Follmerschied}, Definition 6.46, second part. See El Karoui and Quenez\cite{Karouiquenezdp}, F\"ollmer and Kramkov\cite{Follmerkramkov} and  Karatzas and Kou\cite{Karatzaskou} for the original motivation in finance. In Subsection, \ref{labelsectionexamplecoherent}, we characterize the minimal stopping time of a coherent time-consistent risk measure in terms of  the minimal robust representation; see Proposition \ref{labelpropositionuppenvelostimes}.
In Subsection \ref{labelsectionrobuststatistics}, we present a brief discussion from the point of view of robust statistics. We discuss the role of the penalty function in the task of monitoring dates of maximal risk: The more exact the penalty function rates the different models, the better  the timing for intervention is.
%*******************************************************************************************************************************************************
\section{Dynamical convex risk measures in $L^p(R)$} \label{labelsectiondynamicalcrm}
%*******************************************************************************************************************************************************

%*******************************************************************************************************************************************************
\subsection{Conditional convex risk measures} \label{labelsectionrvccrm}
%*******************************************************************************************************************************************************
Measures of risk were introduced in the seminal paper Artzner et al\cite{Artznerdelbaeneberheath}. Robust numerical representations of  risk measures in a general probability space  were obtained by Delbaen\cite{Delbaencmfgps} in the coherent case and extended to the convex case by F\"ollmer and Schied \cite{Follmerschiedcmrrpre,Follmerschiedcmrtradcons} and Fritelli and Rosazza Gianin\cite{Fritelligianin}. Quantifying risk beyond $L^{\infty}(R)$ is the subject of recent research; see e.g., Biagini and Fritelli\cite{Biaginifritelli}, Cheridito and Li\cite{Cheriditoli}, Filipovi\'c and Svindland\cite{Filipovicsvindland}, Kaina and R\"uschendorf\cite{Kainaruschendorf}, Kr\"atschmer\cite{Kratschmer}, Ruszczy\'nski and Shapiro\cite{Ruszczynskishapiro}. Robust numerical representations of conditional convex risk measures in $L^{\infty}(\mf,R)$ are discussed by several authors; see e.g., Bion-Nadal\cite{Bionnadalconditional}, Theorem 3,  and Detlefsen and Scandolo\cite{Detlefsenscandolo}, Theorem 1. A numerical representation for conditional convex risk measures of bounded stochastic processes in discrete time is obtained by Cheridito et al\cite{Cheriditodelbaenkupperdmrmbdtp}, Theorem 3.16. \\

In this section we discuss a robust numerical representation of a real-valued conditional convex risk measure defined in $L^p(\mf,R)$; see Theorem \ref{labelcrobustrepresentationlocallyequivalent} below. This result provides a bridge which connects two main streams in the literature: Real-valued convex risk measures in  $L^{p}(R)$ and conditional convex risk measures in $L^{\infty}(\mf,R)$.\\

Let us introduce some notation. We fix a \textit{complete} probability space $(\Omega, \mf, R)$ and a sub-$\sigma$-algebra $\mg \subset \mf$. We assume that $\mg$ contains the null events of $R$. We fix an exponent $p$ with $1 \leq p < \infty$ and denote by $q$ the conjugate exponent. Typically,  we write $Z^Q$ to denote the density of an absolutely continuous probability measure $Q$. We denote by $\overline{L}^0(\mg,R)$ the family of $\mg$-measurable functions with values in $\mbr\cup\{-\infty, +\infty\}$.\\
 
In the next definition, relationships of random variables hold  $R$-a.s. true.
\fnd{labeldefinitionconditionalconvexriskmeasure}
{
A  conditional convex risk measure $\rho$ in $L^p(\mf,R)$ is a mapping   $\rho:L^p(\mf,R) \to \overline{L}^0(\mg,R)$ with the following properties. For all $X,Y \in L^p(\mf,R)$: 
\begin{enumerate}
	\item Conditional cash invariance: For all $Z \in L^p(\mg,R)$ $\rho(X+Z)=\rho(X)-Z$.
	\item Monotonicity: If $X \leq Y$ $R$-as. then $\rho(X) \geq \rho(Y)$. 
	\item Conditional convexity: For all $\lambda \in L^p(\mg,R)$  with $ 0\leq \lambda \leq 1$ $R$-a.s.:
	\[
	\rho(\lambda X + (1-\lambda)Y) \leq \lambda\rho(X) + (1-\lambda)\rho(Y).
	\]
\end{enumerate}
}
In this section, we fix a conditional convex risk measure $\rho$, which is furthermore normalized:
\eq{labelequationnormalization}
{
\rho(0)=0.
}
The axiomatic framework of Definition \ref{labeldefinitionconditionalconvexriskmeasure} is considered by F\"ollmer and Penner\cite{Follmerpenner} and Detlefsen and Scandolo\cite{Detlefsenscandolo}. Variants of this formulation are considered, e.g.,  by Cheridito et al\cite{Cheriditodelbaenkupperdmrmbdtp} and Weber\cite{Weberdirm}.\\

An important class of conditional convex risk measures are those with the  representability property of Definition  \ref{labeldefinitionrepresentabilidad} below. We need to establish a convention for the conditional expectation of a probability measure which is only absolutely continuous with respect to $R$. Let $Q$ be an absolutely continuous probability measure with density $Z^Q$. For $X \in L^{p}(\mf,R)$, we are going to chose a specific version of the conditional expectation as follows:
\begin{align} \label{labelalignconventionconditionalexpectation}
E_Q[X \mid \mg] &:=  \left\{ 
\begin{array}
{ll}
\frac{1}{E_R[Z^Q \mid \mg]}E_R[Z^Q X \mid \mg], &\mbox{ in } \{E_R[Z^Q \mid \mg]>0\}, \\ 
0, &\mbox{ in } \{E_R[Z^Q \mid \mg]=0\}.
\end{array} 
\right.
\end{align}
With this convention, the essential supremum in  Equation \eqref{labelequationrepresentation} of the next definition,  is unambiguous. We are going to distinguish a special class of absolutely continuous probability measures:
\eq{labelequationabsocontinuouwithdualdensities}
{
\mq^{q}:=\left\{Q \ll R \mid \frac{dQ}{dR} \in L^q(R)\right\}.
}
\nd{labeldefinitionrepresentabilidad}
{
Let $\mq \subset \mq^{q}$ be a class of absolutely continuous probability measures. A penalty function is a correspondence of the form $\alpha:\mq \to \overline{L}^0(\mg,R)$.  The pair $(\mq,\alpha)$ represents the convex risk measure $\rho$ if 
\eq{labelequationrepresentation}
{
\rho(X)= \esssup_{Q \in \mq}\left\{E_Q[-X \mid \mg] - \alpha(Q)\right\}, R-a.s, \mbox{ for  each } X \in L^p(\mf,R).
}
In this case, we say that the conditional convex risk measure $\rho$ is representable and \eqref{labelequationrepresentation} defines a robust representation.
}

As we are going to see in Theorem \ref{labelcrobustrepresentationlocallyequivalent},  conditional risk measures are representable, if the risk measure satisfies  the following regularity condition:
\nd{labeldefinitionfatoupro}
{
The convex risk measure $\rho$ has the Fatou property if
\[
\rho(X)\leq \liminf_{n \to \infty}\rho(X_n), \thinspace R-a.s.,
\]
for each sequence $\{X_n\}_{n=1}^{\infty} \subset  L^p(\mf,R)$ dominated by some $Y \in L^p(\mf,R)$ and converging to $X\in L^p(\mf,R)$.
}
The numerical representation of Theorem \ref{labelcrobustrepresentationlocallyequivalent} below involves the acceptance set and the minimal penalty function associated to $\rho$ in the next definition.
\nd{labeldefinitionminimalpenaltyfunction}
{
The acceptance set of the risk measure $\rho$ is defined by
\[
\ma:= \{a \in L^p(\mf,R) \mid \rho(a) \leq 0, \thinspace R-a.s.\}.
\]
The minimal penalty function 
\[
\alpha^{\min}: \mq^q \to \overline{L}^0(\mg,R)
\] 
is given by
\begin{align} \label{labelequationminimalpenaltyf}
\alpha^{\min}(Q):=  \left\{ 
\begin{array}
{ll}
\esssup_{a \in \ma} \left\{E_Q[-a \mid \mg]\right\}, &\mbox{ in } \{E_R[Z^Q \mid \mg]>0\}, \\ 
+\infty, &\mbox{ in } \{E_R[Z^Q \mid \mg]=0\}.
\end{array} 
\right. 
\end{align}
}
At this point we cannot discard the case where $\alpha^{\min}(Q)$ may be infinite with $Q$-positive probability, but strictly less than one,  for some $Q \in \mq$. Indeed:
\eq{labeleqautioninfinitepenaliyationonnull}
{
\{E_R[Z^Q \mid \mg]=0\} \subset \{\alpha^{\min}(Q)=\infty\}.
}
With this in mind,  we introduce a special  subclass of $\mq^q$:
\eq{labelequationqqinf}
{
\mq^{q,\infty}:=\{Q \in \mq^{q} \mid \alpha^{\min}(Q) \in L^{\infty}(\mg,Q)\}.
}

The next theorem is the main representation theorem of this section. We need to consider the following class of ``locally equivalent'' probability  measures:
\eq{labelequationlocallyequivalentprobabmeas}
{
\mq^{q,\infty}_{e,loc}:=\{Q \in \mq^{q,\infty} \mid E_R[Z^Q \mid \mg]>0, \thinspace R-a.s.\}.
}

\nt{labelcrobustrepresentationlocallyequivalent}
{
If the conditional convex risk measure $\rho$ has the Fatou property, then the  pair $(\mq^{q,\infty}_{e,loc},\alpha^{\min})$ represents the risk measure $\rho$:
\eq{labelequationrobustrepresentation}
{
\rho(X)= \esssup_{Q \in \mq^{q,\infty}_{e,loc}}\left\{E_Q[-X \mid \mg] - \alpha^{\min}(Q)\right\}.
}
}
%******************************************************************************************************************************************************
\subsubsection{Proof of Theorem \ref{labelcrobustrepresentationlocallyequivalent}}
%******************************************************************************************************************************************************
The proof of Theorem \ref{labelcrobustrepresentationlocallyequivalent} needs some preparation. The first step is provided by Proposition \ref{labeltheoremrobustrepresntation}. It gives a ``coarse'' representation of the risk measure $\rho$ in terms of the class $\mq^{q,\infty}$. Then, Lemma \ref{labellemmahalmossavageargumentfornonempt} allow us to refine the representation in terms of the smaller class $\mq^{q,\infty}_{e,loc}$.\\

We start with some remarks and the local property of Lemma \ref{labellemmalocalproperty}.
\nrem{labelremarkacceptanceset}
{
The acceptance set $\ma$ is a convex set with the following properties:
\begin{enumerate}
	\item  It is solid. If $X \in L^p(\mf,R)$, $Y \in \ma$ and $X \geq Y$, then $Y \in \ma$, due to the monotonicity property of the convex risk measure $\rho$. 
\item	If $\rho$ has the Fatou property of Definition  \ref{labeldefinitionfatoupro}, then  $\ma$ is sequentially closed with respect to $R$-a.s. convergence.
\item If $X \in \ma$ and $B \in \mg$, then $1_B X \in \ma$, due to the localization property of  Lemma \ref{labellemmalocalproperty} below.
\end{enumerate}
}
\nrem{labelremarkanothercharmpf}
{
The minimal penalty function can equivalently be defined by
\[
\alpha^{\min}(Q)= \esssup_{X \in L^p(\mf,R)} \left\{E_Q[-X \mid \mg]- \rho(X)\right\}.
\]
}

\nl{labellemmalocalproperty}
{
A real-valued conditional convex risk measure $\rho$ has the following localization property. For each $A \in \mg$:
\[
\rho(1_A X + 1_{A^c}Y) = 1_A \rho(X) + 1_{A^c}\rho(Y).
\]
}
Proof. This property follows from the property of conditional convexity; see Detlefsen and Scandolo\cite{Detlefsenscandolo}, Proposition 1.\finlemma\\

The next proposition provides a ``coarse'' representation. The proof applies the Hahn-Banach hyperplane separating theorem and follows the original ideas of \cite{Delbaencmfgps,Follmerschiedcmrrpre,Follmerschiedcmrtradcons,Fritelligianin}.
\np{labeltheoremrobustrepresntation}
{
Let $\rho$ be a  conditional convex risk measure in $L^p(\mf,R)$. If $\rho$ has the Fatou property, then for each $X \in L^p(\mf,R)$:
\eq{labelequationrobustrepresentationcoarse}
{
\rho(X)= \esssup_{Q \in \mq^{q,\infty}}\left\{E_Q[-X \mid \mg] - \alpha^{\min}(Q)\right\}, \thinspace R-a.s.
}
}
Proof. 
\begin{enumerate}
	\item Let $X \in L^p(\mf,R)$. We set
\eq{labelequationdefb}
{
b:=\esssup_{Q \in \mq^{q,\infty}}\left\{E_Q[-X \mid \mg] - \alpha^{\min}(Q)\right\}.
} 
We must show that $R[\rho(X) = b]=1$. It is clear that  $R[\rho(X) \geq b]=1$, due to the definition of the minimal penalty function.  Now we show the converse inequality. Assume by way of contradiction that 
\[
R[\rho(X) > b]>0.
\]
Let us call 
\[
J:=\{\rho(X) > b\}.
\]
Note that $J \in \mg$ and 
\[
\rho(X) - b = 1_J(\rho(X) - b).
\]
Moreover,
\[
1_J(\rho(X) - b)= \rho(1_J(X+b)).
\]
Thus, $1_J(X+b)$ does not belong to the acceptance set $\ma$. 
\item Now we separate the sets $\ma$ and $\{1_{J}(X+b)\}$. There exists a linear functional $l:L^p(\mf,R) \to \mbr$ such that
\begin{align}
\inf_{a \in \ma} l(a) &\geq x,\label{labelequationseparating}\\
 l(1_{J}(X+b))&<x, \label{labelequationseparatingsecond}
\end{align}
due to the Hahn-Banach hyperplane separating Theorem; see e.g., F\"ollmer and Schied\cite{Follmerschied}, Theorem A.56. Note that  $x\leq 0$, since $0 \in \ma$ and $l(0)=0$.
\item The linear functional $l$ can be selected to be of the form
\[
l(X)=E_{Q^0}[X], \mbox{ for each } X \in L^p(\mf,R),
\]
where the probability measure $Q^0$ is absolutely continuous with respect to $R$ and the density $\frac{dQ^0}{dR}$ belongs to $L^q(\mf,R)$. Indeed, this follows from the fact that  $\ma$ is a solid convex set; see Remark \ref{labelremarkacceptanceset}, first part, and the Riesz representation Theorem of linear functionals of $L^p(\mf, R)$.
\item The inequality \eqref{labelequationseparating} implies
\eq{labeleuqationprimerconsecuencia}
{
E_{Q^0}[a \mid \mg] \geq x, \thinspace Q^0-a.s.,  \mbox{ for each } a \in \ma.
}
Indeed, for $a \in \ma$, the random variable 
\[
\widehat{a}:= a 1_{\{E_{Q^0}[a \mid \mg] < x\}}
\]
belongs to the acceptance set $\ma$, since $\{E_{Q^0}[a \mid \mg] < x\} \in \mg$; see Remark \ref{labelremarkacceptanceset}, third part. Thus, $E_{Q^0}[\widehat{a}] \geq x$. On the other hand, 
\[
E_{Q^0}[\widehat{a}]= E_{Q^0}[E_{Q^0}[\widehat{a} \mid \mg]]= E_{Q^0}[1_{\{E_{Q^0}[a \mid \mg] < x\}} E_{Q^0}[a  \mid \mg]] \leq x.
\]
Thus, $Q^0[\{E_{Q^0}[a \mid \mg] < x\}]=0$ and \eqref{labeleuqationprimerconsecuencia} holds true.\\

Note that  $\alpha^{\min}(Q^0) \leq -x$, $Q^0-a.s.$, due to \eqref{labeleuqationprimerconsecuencia}. Hence, $Q^0 \in \mq^{q,\infty}$.\\

\item Now let us define
\[
J':= \{E_{Q^0}[1_J(X+b)\mid \mg] < x\}.
\]
Then, $J' \in \mg$ and $J' \subset J$. Moreover 
\[
{Q^0}[J'] >0,
\]
due to the inequality \eqref{labelequationseparatingsecond}. 
\item Now we generate a contradiction.  In the event $J'$ we have
\[
\alpha^{\min}({Q^0}) < E_{Q^0}[-1_J(X+b)\mid \mg], \thinspace {Q^0}-a.s.,
\]
due to the definitions of the minimal penalty function and of the event $J'$.  We may rewrite this last inequality to obtain
\[
 b < E_{Q^0}[-X\mid \mg]-  \alpha^{\min}({Q^0}), \thinspace Q^0-a.s. \mbox{ in the event } J'. 
\]
This contradicts the definition of $b$ given in \eqref{labelequationdefb}.\fintheo
\end{enumerate}
To some extend, it is unpleasant to select a specific version of the conditional expectation, as fixed in  \eqref{labelalignconventionconditionalexpectation}. The convention is unnecessary for probability measures $Q \ll R$ with
\[
E_R[Z^Q \mid \mg]>0, \thinspace R-a.s.
\]
In the next lemma we show that the class of ``locally equivalent'' probability measures \eqref{labelequationlocallyequivalentprobabmeas}
\[
\mq^{q,\infty}_{e,loc}:=\{Q \in \mq^{q,\infty} \mid E_R[Z^Q \mid \mg]>0, \thinspace R-a.s.\},
\]
is non empty. We use an exhaustion argument due to  Halmos and Savage. The exhaustion argument in the theory of risk measure is well known; see e.g.,  Cheridito et al\cite{Cheriditodelbaenkupperdmrmbdtp}, Lemma 3.22 and F\"ollmer and Penner\cite{Follmerpenner}, Lemma 3.4. 
\nl{labellemmahalmossavageargumentfornonempt}
{
If $\rho$ has the Fatou property, then the class $\mq^{q,\infty}_{e,loc}$ is non empty. 
}
Proof.
\begin{enumerate}
\item We define 
\[
c:= \sup\left\{R(E_R[Z^Q \mid \mg] >0) \mid Q \in \mq^{q,\infty}\right\}.
\]	
There exists $Q^* \in \mq^{q,\infty}$ such that 
\[
c= R(E_R[Z^{Q^*} \mid \mg] >0). 
\]
Indeed, let $Q^n$ be a maximizing sequence, so that 
\[
c= \lim_{n \to \infty} R(E_R[Z^{Q^n} \mid \mg] >0).
\]
We define 
\[
\lambda^n:=\frac{1}{2^n}\frac{1}{1+ \left\|Z^n\right\|^q_{L^{q}(R)} + \left\|\alpha^{\min}(Q^n)\right\|_{L^{\infty}(Q^n)}}.
\]
Then, the probability measure $Q^* \ll R$ defined by the density
\[
\frac{dQ^*}{dR}:= \frac{1}{E_R\left[\sum_{n=1}^{\infty} \lambda^n \frac{dQ^n}{dR}\right]}\sum_{n=1}^{\infty} \lambda^n \frac{dQ^n}{dR}
\]
is an element of $\mq^{q,\infty}$. It  attains the value $c$:
\[
c= R(E_R[Z^{Q^*} \mid \mg] >0).
\]
\item Let $A \in \mg$ with $R(A)>0$. Then we have
\[
1_A = \rho(-1_A)= \esssup_{Q \in \mq^{q,\infty}}\left\{E_Q[1_A \mid \mg] - \alpha^{\min}(Q)\right\},
\]
due to Proposition \ref{labeltheoremrobustrepresntation}. Hence, we conclude the existence of $\widehat{Q} \in \mq^{q,\infty}$ with
\[
\left\{E_R[Z^{\widehat{Q}}\mid \mg]>0\right\} \cap A \neq \emptyset, R-a.s.,
\]
since $\rho$ is normalized and our convention of the conditional expectation \eqref{labelalignconventionconditionalexpectation}.
\item Now we conclude the proof by showing that $c=1$. Assume by way of contradiction that $c<1$. Let the event $A \in \mg$ be defined by 
\[
A:=\left\{E_R[Z^{Q^*} \mid \mg] =0\right\}.
\]
There exists $\widehat{Q} \in \mq^{q,\infty}$ with
\[
\left\{E_R[Z^{\widehat{Q}}\mid \mg]>0\right\} \cap A \neq \emptyset, R-a.s.,
\]
due to the previous step. The probability measure defined by
\[
Q^0:=\frac{1}{2}(Q^*+\widehat{Q}),
\]
belongs to the class $\mq^{q,\infty}$. It contradicts the optimality of $Q^*$ since
\[
\left\{E_R[Z^{Q^0} \mid \mg] >0\right\}=\left\{E_R[Z^{Q^*} \mid \mg] >0\right\} \cup \left\{E_R[Z^{\widehat{Q}} \mid \mg] >0\right\}.\fintheo
\]
\end{enumerate}

Now we are ready to prove Theorem \ref{labelcrobustrepresentationlocallyequivalent}.\\
Proof.
Let $X\in L^p(\mf,R)$ and $Q^0 \in \mq^{q,\infty}$. The identity  \eqref{labelequationrobustrepresentation}  will be established after we construct a probability measure  $\widetilde{Q} \in \mq^{q,\infty}_{e,loc}$ such that
\eq{labelequationafproving}
{
E_{Q^0}[-X \mid \mg] - \alpha^{\min}(Q^0) \leq E_{\widetilde{Q}}[-X \mid \mg] - \alpha^{\min}(\widetilde{Q}), \thinspace R-a.s,
}
due to Proposition \ref{labeltheoremrobustrepresntation}. 
We set $A:=\{E_R[Z^{Q^0} \mid \mg]=0\}$. Assume $0<R(A)<1$, otherwise there is nothing to prove.\\

The class $\mq^{q,\infty}_{e,loc}$ is non empty, due to Lemma \ref{labellemmahalmossavageargumentfornonempt}. Without loss of generality, we assume that $R \in \mq^{q,\infty}_{e,loc}$. We define a probability measure $\widetilde{Q} \ll R$ by
\[
\frac{d\widetilde{Q}}{dR}:=Y 1_A + Z^{Q^0}1_{A^c}.
\]
In this expression, $Y$ is a positive constant selected to satisfy $E_R\left[\frac{d\widetilde{Q}}{dR}\right]=1$. The probability measure $\widetilde{Q}$ belongs to $\mq^{q,\infty}_{e,loc}$ by construction.\\

The penalization and conditional expectation of  $\widetilde{Q}$ can be computed as follows:
\begin{align} \notag
\alpha^{\min}(\widetilde{Q})&= 1_A \alpha^{\min}(\widetilde{Q}) + 1_{A^c}\alpha^{\min}(Q^0), \notag\\
E_{\widetilde{Q}}[-X \mid \mg]
&= \frac{1_A}{Y} E_{R}[-YX \mid \mg] +  \frac{1_{A^c}}{E_{R}[Z^{Q^0} \mid \mg]} E_{R}[-Z^{Q^0}X \mid \mg] \notag\\
&= 1_A E_{R}[-X \mid \mg] +  1_{A^c} E_{Q^0}[-X \mid \mg]. \notag
\end{align}
Thus, \eqref{labelequationafproving} holds true, due to the set relationship \eqref{labeleqautioninfinitepenaliyationonnull}.\fincor
%*******************************************************************************************************************************************************
\subsection{Real-valued conditional convex risk measures}\label{labelsectionrealvaluedccrm}
%*******************************************************************************************************************************************************
\nd{labeldefinitionrealvalueconditionalcrm}
{
A conditional convex risk measure $\rho:L^p(\mf,R) \to \overline{L}^0(\mg,R)$ is real-valued if it takes values in $L^1(\mg,R)$. More precisely, for each $X \in L^p(\mf,R)$ we have $\rho(X) \in L^1(\mg,R)$.
}
\nl{labellemmacontinuity}
{
Let $\rho$ be a real-valued conditional convex risk measure in $L^p(\mf,R)$. Let $\{X^n\}_{n=1}^{\infty} \subset L^p(\mf,R)$  be a sequence strongly converging to $X^0 \in L^p(\mf,R)$. Then 
\[
\lim_{n \to \infty} \left\|\rho(X^n)-\rho(X^0)\right\|_{L^1}=0.
\]
}
Proof.
We start with $X^0=0$. We assume that 
\[
\lim_{n \to \infty} 2^n  \left\|X^n\right\|_{L^p} =0,
\]
by taking a subsequence if necessary. The sequence defined by
\[
Y^n:= \sum_{i=1}^{n} \frac{1}{2^j \left\|X^j\right\|_{L^p}} \left|X^j\right|
\]
is increasing. It is easy to see that the sequence has the Cauchy property. Thus, it converges to some $Y \in L^p(\mf,R)$.\\

Now we get
\[
\left|\rho(X^n)\right| \leq \rho(-\left|X^n\right|) \leq  2^n \left\|X^n\right\|_{L^p}   \rho(-\frac{1}{2^n \left\|X^n\right\|_{L^p}}\left|X^n\right|),
\]
due to monotonicity and convexity of $\rho$. Hence
\[
\left|\rho(X^n)\right| \leq 2^n \left\|X^n\right\|_{L^p}   \rho(-Y).
\]

In order to obtain the result for arbitrary $X^0$, we define a new convex risk measure by $\rho^0(X):=\rho(X + X^0)- \rho(X^0)$. It is easy to see that $\rho^0$ satisfies the conditions of the proposition. Thus, we may apply the previous step.\finpro\\

\nrem{labelremarkmimicnamioka}
{
The arguments in the  proof of Lemma \ref{labellemmacontinuity} are due to  Biagini and Fritelli\cite{Biaginifritelli}, Theorem 2 (which they call  extended Namioka-Klee Theorem). Note that Lemma \ref{labellemmacontinuity} does not follow directly from Theorem 2 \cite{Biaginifritelli} by considering the functional $E_R[\rho]$. 
}
\nc{labelcorollaryfatoupropertyautomaticforrealvalued}
{
A real-valued convex risk measure has the Fatou property.
}

\nt{labeltheoremrepresentabilityofrealvaluedccrm}
{
A real-valued conditional convex risk measure is representable.
}
Proof. Any real-valued conditional convex risk measure has the Fatou property due to Corollary \ref{labelcorollaryfatoupropertyautomaticforrealvalued}. Thus, representability holds true due to Theorem \ref{labelcrobustrepresentationlocallyequivalent}.\fintheo  
%*******************************************************************************************************************************************************
\subsubsection{Coercivity}
%*******************************************************************************************************************************************************
\providecommand{\cnorm}[1]{\left\lceil  #1\right\rceil}
In Theorem \ref{labeltheoremrepresentabilityofrealvaluedccrm} we have seen that real-valued conditional risk measures are representable. In Theorem \ref{labeltheoremcoercivitypropertynecessity} below, we are going to see that the  penalty functions of  real-valued convex risk measures  must satisfy the coercivity property \eqref{labelequationcoercivitypropertydef} below. Let us introduce the class
\begin{align}
\mq^{q}_{e,loc}&:=\{Q \in \mq^{q} \mid E_R[Z^Q\mid \mg]>0, R-a.s.\}.\notag
\end{align}
The main concept of this section is that of coerciveness. This concept is introduced by Cheridito and Li\cite{Cheriditoli}, Definition 4.6, in the non-conditional case.
\nd{labeldefinitioncoercivitypf}
{
Let $\alpha: \mq^{q}_{e,loc} \to \overline{L}^0_+(\mg,R)$ be a penalty function. We say that $\alpha$ is a coercive penalty function if there exist real constants $a,b$ with $b>0$ such that
\eq{labelequationcoercivitypropertydef}
{
E_R[\alpha(Q)] \geq a + b \thinspace E_R\left[\frac{1}{E_R\left[\frac{dQ}{dR} \mid \mg\right]}E^{\frac{1}{q}}_R\left[\left(\frac{dQ}{dR}\right)^q\mid \mg\right]\right], \thinspace Q \in \mq^{q}_{e,loc}.
}

}
The first result of this section is Proposition \ref{labelpropositionsufficiencycoercivity}. In  this proposition we show that coercive penalty functions define real-valued conditional risk measures in $L^p(\mf,R)$. To prove this result, we need some preparation.\\

We are going to denote by $L^p_+(\mf,R)$ the non negative elements of $L^p(\mf,R)$. We need to introduce the following family of random variables:
\[
S^{+}:=\{X\in L^p_+(\mf,R) \mid  \left\|X\right\|_{L^p}=1\}.
\]
The second part of the next lemma uses a well known argument about the linear functionals of the space $L^p(\mf,R)$; see e.g., Werner\cite{Wernerfunktionalanalysis}, Beispiel (j), p.50.
\nl{labellemmacoercividad}
{
Let  $Z \in L^q_+(\mf,R)$. Then the random variable $\cnorm{Z}$ defined by
\eq{labelequationconditionalnorm}
{
\cnorm{Z}:= \esssup_{X \in S^{+}}E_R[ZX\mid \mg],
}
belongs to $L^{1}(\mg,R)$. Moreover
\eq{labelequationexplicitanorma}
{
\cnorm{Z}=E^{\frac{1}{q}}_R[Z^q \mid \mg].
}
}
Proof. 
\begin{enumerate}
	\item Let $\{Y_n\}_{n=1}^{\infty} \subset S^{+}$ be a maximizing sequence: 
\[
\lim_{n\to \infty} E_R[ZY_n\mid \mg]=\cnorm{Z}.
\]	
Then we get
\[
0 \leq E_R[\cnorm{Z}] \leq \liminf_{n\to \infty}E_R[ZY_n],
\]
due to Fatou's Lemma.  Moreover,
\[
E_R[ZY_n] \leq \left\|Z\right\|_{L^q}\left\|Y_n\right\|_{L^p}=\left\|Z\right\|_{L^q},
\]
due to H\"older's inequality. Thus:
\[
0 \leq E_R[\cnorm{Z}] \leq \left\|Z\right\|_{L^q}.\\
\]
\item Now we prove \eqref{labelequationexplicitanorma}. We define
\[
X^0:=\frac{Z^{\frac{q}{p}}}{E^{\frac{1}{p}}_R[Z^q\mid \mg]}1_{\{E_R[Z\mid \mg] >0\}}.
\]
The random variable $X^0$ is well defined and belongs to $S^+$, since 
\[
\{E_R[Z^q\mid\mg]=0\} \subset \{Z^q=0\}.
\]
Moreover
\[
E_R[X^0 Z \mid \mg]=E^{\frac{1}{q}}_R[Z^q \mid \mg].
\]
Hence
\[
\cnorm{Z} \geq E^{\frac{1}{q}}_R[Z^q \mid \mg].\finlemma
\]
\end{enumerate}

\nl{lemmaconditionalholder}
{
Let  $Z \in L^q_+(\mf,R)$ and   $X \in L^p_+(\mf,R)$. Then
\eq{labelequationconditionalholder}
{
E_R[ZX \mid \mg] \leq \cnorm{Z} \left\|X\right\|_{L^p}.
}
}
Proof. Without loss of generality we may assume that $\left\|X\right\|_{L^p}>0$. The following relationship is clear
\[
E_R[ZX \mid \mg]=E_R\left[Z \frac{X}{\left\|X\right\|_{L^p}} \mid \mg\right]\left\|X\right\|_{L^p}.
\]
Hence
\[
E_R[ZX \mid \mg] \leq \cnorm{Z}\left\|X\right\|_{L^p}.\finlemma
\]

After this preparation we are now ready to prove that  coercive penalty functions induce real-valued risk measures. The next proposition is the conditional version of the first part of Proposition  4.7 of Cheridito and Li\cite{Cheriditoli} and  the first part of Proposition 2.10 of Kaina and R\"uschendorf\cite{Kainaruschendorf}.
\np{labelpropositionsufficiencycoercivity}
{
Let $\alpha: \mq^{q}_{e,loc} \to \overline{L}^0_+(\mg,R)$ be a  penalty function.  Assume there exists $Q^0 \in \mq^{q}_{e,loc}$ such that $\alpha(Q^0) \in L^{\infty}(\mg,R)$.  Let us define a mapping $\rho$ by 
\[
\rho(X):= \esssup_{Q \in \mq^{q}_{e,loc}}\left\{E_Q[-X \mid \mg] - \alpha(Q)\right\}, X \in L^p(\mf,R).
\]
Then $\rho$ is a real-valued conditional risk measure, if $\alpha$ is coercive.
}
Proof. We only prove that $\rho$ is real-valued. Note that it is only necessary to prove the result for the minimal penalty function $\alpha^{\min}$.
\begin{enumerate}
\item We first consider the case where $X$ is non positive. There exists $\widetilde{X} \in L^{\infty}$ such that $X \leq \widetilde{X} \leq 0$ and
\[
\left\|X-\widetilde{X}\right\|_{L^p} \leq \frac{b}{2},
\]
due to Lebesgue Dominated Convergence Theorem.  Hence
\begin{align}
\rho(X)&=\esssup_{Q \in \mq}\left\{E_Q[-\widetilde{X} \mid \mg] +  E_Q[\widetilde{X} -X\mid \mg] - \alpha^{\min}(Q)\right\} \notag\\
&\leq \left\|\widetilde{X}\right\|_{L^{\infty}}  +  \esssup_{Q \in \mq}\left\{E_Q[\widetilde{X} -X\mid \mg] -  \alpha^{\min}(Q)\right\}. \label{labelequationprimeraestimacionpncoer}
\end{align}
\item If we take expectation with respect to $R$, then we get
\eq{labelequationlatticeimplica}
{
E_R\left[\esssup_{Q \in \mq}\left\{E_Q[\widetilde{X} -X\mid \mg] -  \alpha^{\min}(Q)\right\}\right]
=
\sup_{Q \in \mq}E_R\left[\left\{E_Q[\widetilde{X} -X\mid \mg] -  \alpha^{\min}(Q)\right\}\right].
}
Indeed, let $\{Q^n\}_{n=1}^{\infty} \subset \mq$ be a maximizing sequence:
\[
\esssup_{Q \in \mq}\left\{E_Q[\widetilde{X} -X\mid \mg] -  \alpha^{\min}(Q)\right\} = \lim_{n\to \infty}\left\{E_{Q^n}[\widetilde{X} -X\mid \mg] -  \alpha(Q^n)\right\}.
\]
We claim that we may assume  the sequence of functions $\left\{E_{Q^n}[\widetilde{X} -X\mid \mg] -  \alpha^{\min}(Q^n)\right\}$ to be  bounded from below. Hence, we can apply Fatou's Lemma to obtain:
\[
E_R\left[\esssup_{Q \in \mq}\left\{E_Q[\widetilde{X} -X\mid \mg] -  \alpha^{\min}(Q)\right\}\right] \leq  \liminf_{n\to \infty}E_R\left[E_{Q^n}[\widetilde{X} -X\mid \mg] -  \alpha^{\min}(Q^n)\right].
\]
This proves the inequality $\leq$. The converse direction is clear.\\

Now we prove the claim. For $Q^n$ we define the event
\[
A^n:=\left\{E_{Q^n}[\widetilde{X} -X\mid \mg] -  \alpha(Q^n) \geq - \left\|\alpha^{\min}(Q^0)\right\|_{L^{\infty}} \right\}.
\]
We construct a probability measure $\widetilde{Q}^n$ by
\[
\frac{d\widetilde{Q}^n}{dR}:= 1_{A^n} \frac{dQ^n}{dR} + 1_{(A^n)^c} \frac{dQ^0}{dR}Y.
\]
The penalization and conditional expectation of  $\widetilde{Q}$ can be computed as follows:
\begin{align} \notag
\alpha^{\min}(\widetilde{Q}^n)&= 1_{A^n} \alpha^{\min}(Q^n) + 1_{(A^n)^c}\alpha^{\min}(Q^0), \notag\\
E_{\widetilde{Q}^n}[-X \mid \mg] &= 1_{A^n} E_{Q^n}[\widetilde{X} -X\mid \mg] +  1_{(A^n)^c} E_{Q^0}[\widetilde{X} -X\mid \mg]. \notag
\end{align}
\item Now we put together  \eqref{labelequationprimeraestimacionpncoer} and\eqref{labelequationlatticeimplica} to obtain
\begin{align}
E_R[\rho(X)] \leq \left\|\widetilde{X}\right\|_{L^{\infty}}  +  \sup_{Q \in \mq} E_R\left[E_Q[\widetilde{X} -X\mid \mg] -  \alpha^{\min}(Q)\right]. \notag
\end{align}
Moreover
\begin{align} \notag
E_R\left[E_Q[\widetilde{X} -X\mid \mg]\right]&=E_R\left[\frac{1}{E_R[Z^Q \mid \mg]}E_R[Z^Q(\widetilde{X}-X)\mid \mg]\right]\notag\\
&\leq E_R\left[\frac{1}{E_R[Z^Q \mid \mg]}\cnorm{Z^Q}\right]\left\|X-\widetilde{X}\right\|_{L^p},\notag
\end{align}
due to Lemma \ref{lemmaconditionalholder}. Thus
\[
\sup_{Q \in \mq}E_R\left[E_Q[\widetilde{X} -X\mid \mg] -  \alpha(Q)\right] \leq -a  -\frac{b}{2} \sup_{Q \in \mq}
E_R\left[\frac{1}{E_R[Z^Q \mid \mg]}\cnorm{Z^Q}\right].  
\]
Hence, we conclude:
\[
E_R[\rho(X)] \leq \left\|\widetilde{X}\right\|_{L^{\infty}} - a. 
\]
\item  Now we consider the case $X \geq 0$.  We have
\[
0 \geq \rho(X) \geq E_{Q^0}[-X \mid \mg] - \alpha^{\min}(Q^0).
\]
The conditional expectation $E_{Q^0}[-X \mid \mg]$ is integrable since $X \in L^p(\mf,R)$ and $\frac{dQ^0}{dR} \in L^q(\mf,R)$. The penalization $\alpha^{\min}(Q^0)$ is integrable by hypothesis.  Hence $\rho(X)$ belongs to $L^1(\mg,R)$.
\item For general $X$ we write $X=X^+ +  X^-$ with $X^+\geq 0$ and $X^- \leq 0$. We have
\[
\rho(X^-) \geq \rho(X) \geq \rho(X^+),
\]
due to the monotonicity of conditional expectation. We conclude the desired integrability of $\rho(X)$ from the previous steps.\finlemma
\end{enumerate}

The previous proposition established that coercive penalty functions induce real-valued conditional risk measures. Now we prove the converse. This is the main result of this section.
\nt{labeltheoremcoercivitypropertynecessity}
{
Let $\rho$ be a real-valued conditional convex risk measure.  If the pair $(\mq^q_{e,loc},\alpha)$ represents the convex risk measure $\rho$, then the penalty function $\alpha$ must be coercive.
}
Proof.  By way of contradiction assume there exists a sequence $\{Q^n\}_{n=1}^{\infty} \subset \mq^q_{e,loc}$ such that 
\eq{labelequationbreakingthecoerc}
{
E_R\left[\alpha(Q^n)\right] <-n + 2^{-n-1}E_R\left[\frac{1}{E_R[Z^n\mid \mg]}\cnorm{Z^n}\right],
}
where $Z^n$ denotes the density of $Q^n$. There exists  $X^n \in S^+$  such that
\[
E_R\left[\frac{E_R[Z^nX^n\mid \mg]}{E_R[Z^n\mid \mg]}\right] \geq \frac{1}{2}E_R\left[\frac{\cnorm{Z^n}}{E_R[Z^n\mid \mg]}\right],
\]
due to Lemma \ref{labellemmacoercividad} and Fatou's Lemma. Now we define
\[
X:=\sum_{n=1}^{\infty}2^{-n}X^n.
\]
We get
\[
\rho(-X)\geq \rho(-2^{-n}X^n)\geq E_{Q^n}[2^{-n}X^n \mid \mg] - \alpha(Q^n).
\]
Moreover
\begin{align} \notag
E_R\left[E_{Q^n}[2^{-n}X^n \mid \mg] - \alpha(Q^n)\right] 
&\geq 
2^{-n-1}E_R\left[\frac{\cnorm{Z^{n}}}{E_R[Z^{n}\mid \mg]}\right] +n -2^{-n-1}E_R\left[\frac{\cnorm{Z^{n}}}{E_R[Z^{n}\mid \mg]}\right]\notag\\
& =n.\notag
\end{align}
Thus:
\[
E_R[\rho(-X)] \geq n, 
\]
a clear contradiction.\fintheo
%****************************************************************************************************************************************************
\subsubsection{Invariance of the minimal representation}
%****************************************************************************************************************************************************
Recall that $\rho$ is a real-valued conditional convex risk measure defined in $L^p(\mf,R)$. We may define a new risk measure $\rho^{\infty}$ in $L^{\infty}(\mf,R)$ by
\[
\rho^{\infty}(X)=\rho(X).
\]
The new risk measure $\rho^{\infty}$ has associated a minimal penalty function:
\eq{labelequationminimalpenaltyfrestricted}
{
\alpha^{\min,\infty}(Q):= \esssup_{X \in L^{\infty}(\mf,R)} \left\{E_Q[-X\mid \mg] - \rho^{\infty}(X)\right\}, \thinspace Q\ll R.
}

In this section we show that the minimal representation is invariant to this restriction in the sense that  $\alpha^{\min,\infty}=\alpha^{\min}$ in $\mq^{q}_{e,loc}$.  
\nt{labelpropositioninvarincia}
{
The minimal representation of $\rho$ keeps invariant in $L^{\infty}(\mf,R)$. Thus,  $\alpha^{\min,\infty}=\alpha^{\min}$ in $\mq^{q}_{e,loc}$.
}
Proof.  Let $Q \in \mq^{q}_{e,loc}$.  Let $Y \in L^p(\mf,R)$. We define a sequence by $Y^n:=(Y \wedge n)\vee(-n)$.  It is clear that $Y^n \in L^{\infty}(\mf,R)$ and the sequence converges to $Y$ in $L^p(\mf,R)$. Moreover, 
\[
\lim_{n \to \infty }E_Q[-Y^n  \mid \mg] = E_Q[-Y \mid \mg], R-a.s.
\]
due to Lebesgue's dominated convergence Theorem and H\"older's inequality. Furthermore,
\[
\lim_{n \to \infty }\rho(Y^n) = \rho(Y),
\]
due to Lemma \ref{labellemmacontinuity}. We conclude that
\[
E_Q[-Y \mid \mg] - \rho(Y) \leq \esssup_{X \in L^{\infty}(\mf,R)}\left\{E_Q[-X \mid \mg] - \rho(X)\right\} = \alpha^{\min,\infty}(Q).
\]
This proves the claims of the theorem.\fintheo

\nrem{labelremarkcharacterization}
{
Theorems \ref{labeltheoremcoercivitypropertynecessity} and \ref{labelpropositioninvarincia} characterize conditional convex risk measures defined in $L^{\infty}(\mf,R)$ which can be extended to  real-valued conditional convex risk measures in $L^p(\mf,R)$.
}
%*******************************************************************************************************************************************************
\subsection{Time consistency} \label{labelsectionrvccrmtimeconsistent}
%*******************************************************************************************************************************************************
Now we introduce a filtration $\mbf:=\{\mf_t\}_{t=0,1,\ldots,T}$  of the probability space $(\Omega, \mf, R)$. The horizon $T$ is finite:  $T<\infty$. We assume that $\mf_T=\mf$ and $\mf_0$ is the $\sigma$-algebra of null events.

\nd{labeldefinitiontimeconsistency}
{
For each $t \in \{0,1,\ldots, T\}$ let  $\rho_t:L^1(\mf_T,R) \to L^1(\mf_{t},R)$ be a conditional convex risk measure. The sequence of conditional risk measures  $\Phi:=\{\rho_t\}_{t=0,\ldots,T}$ is a \textrm{time-consistent} dynamical risk measure if for each $t \in \{0,\ldots, T-1\}$:
\eq{labelequationrecursiveness}
{
\rho_t=\rho_{t} \circ (-\rho_{t+1}).
}
}
Dynamical convex risk measures have been intensively studied; see  e.g., Delbaen\cite{Delbaenastablesets}, Detlefsen and Scandolo\cite{Detlefsenscandolo}, F\"ollmer and Penner\cite{Follmerpenner}, Riedel\cite{Riedeldcrm},  Tutsch\cite{Tutschdrm}, Wang\cite{Wangdynamic}, Weber \cite{Weberdirm}.\\

In particular our axiomatic framework, as  given by Definitions \ref{labeldefinitionconditionalconvexriskmeasure} and \ref{labeldefinitiontimeconsistency}, is consistent with  \cite{Follmerpenner}. Thus, the characterization of time-consistency in terms of the minimal robust representation of F\"ollmer and Penner\cite{Follmerpenner}, Theorem 4.5, can be extended to a version in $L^1(R)$, if the minimal penalty function is coercive in $L^1(R)$.
%*******************************************************************************************************************************************************
\section{Stopping times of maximal risk} \label{labelsectionexistenceofopst}
%*******************************************************************************************************************************************************
In this section we start the second part of the paper.  Recall we fixed a probability space $(\Omega, \mf,R)$ and a filtration $\mbf$ of the space.  We furthermore fix a time-consistent convex risk measure $\Phi=\{\rho_t\}_{t=0,\ldots,T}$ defined in $L^1(\mf,R)$. By $\ftp$ we denote the family of stopping times of the filtration $\mbf$.\\

The motivation of this part is the monitoring of a system and  determining the best time to intervene.  The system will be represented by a non negative stochastic process  $H:=\{H_t\}_{t=0,\ldots,T}$, satisfying 
\[
E_R[H_t] <\infty \mbox{ for each } t=0,\ldots, T.
\]
The ``measuring device'' is the risk measure $\Phi$. The process $H$ may have different interpretations. The idea of monitoring through a risk measure goes back to the seminal paper Artzner et al\cite{Artznerdelbaeneberheath}. We focus on monitoring over time, and thus, as we motivated in the introduction, the maximal risk takes the form
\[
\sup_{\theta \in \ftp}\rho_0(H_{\theta}).
\]
Hence, a stopping time $\tp^* \in \ftp$ attaining this supremum is of special interest. In this section, Theorem \ref{rcace}, we show that such optimal stopping times exists due to the property of time-consistency of the risk measure $\Phi$.\\

It will be necessary to consider starting points other than zero.

\nd{labeldefinitionstomr}
{
The upper Snell envelope of $H$ with respect to $\Phi$ is the stochastic process defined by
\eq{labelequationuppersnellenvelope}
{
U^{\uparrow}_t:= \esssup_{\theta \geq t} \rho_t(-H_{\theta}).
}

A stopping time $\tau_t$ is of $t$-maximal risk if
\[
\rho_t(-H_{\tau_t})= \us.
\]
}
Upper Snell envelopes in the context of American-option pricing is systematically studied by  F\"ollmer and Schied\cite{Follmerschied}, Section 6.5. In continuous time this concept has been considered by  Delbaen\cite{Delbaenastablesets}, El Karoui and Quenez\cite{Karouiquenezdp}, F\"ollmer and Kramkov\cite{Follmerkramkov} and Karatzas and Kou\cite{Karatzaskou}.  The next lemma is a convex version of Theorem 6.52 in  \cite{Follmerschied}.
\nl{labellemmarecurusenvelop}
{
The upper Snell envelope has the following properties:
\begin{enumerate}
	\item It dominates from above the payoff $H$:
\eq{labelequationdomin}
{
H_t \leq U^{\uparrow}_t \thinspace R-a.s.
}
Equality holds $R$-a.s. for $t=T$.
\item  It can be computed recursively as follows:
\eq{labelequationbwcomputation}
{
U^{\uparrow}_t= H_t \vee \rho_t(-U^{\uparrow}_{t+1}).
}
\end{enumerate}
}
Proof. The first claim of the lemma holds true due to the normalization property of $\Phi$.\\

Now we prove the second claim. Let $\theta$ be a stopping time with $t \leq \theta \leq T$. Then
\[
\rho_t(-H_{\theta}) = 1_{\{\theta=t\}} H_t + 1_{\{\theta>t\}} \rho_t(-H_{\theta}),
\]
due to the localization property of Lemma \ref{labellemmalocalproperty}. This identity clearly implies the following inequality:
\[
\rho_t(-H_{\theta}) \leq   H_t \vee  \rho_t(-\us[t+1]).
\]
Thus, the inequality $\leq$ in \eqref{labelequationbwcomputation} holds true. Now we prove the converse.  There exists a sequence of stopping times $\{\theta^n\}_{n=1}^{\infty}$ with $t+1 \leq \theta^n \leq T$ such that
\[
\rho_{t+1}(-H_{\theta^n}) \to \us[t+1].
\]
Then:
\[
\us \geq \rho_{t}(-H_{\theta^n}) = \rho_t(-\rho_{t+1}(-H_{\theta^n})),
\]
due to the time-consistency of the risk measure $\rho$. Hence:
\[
\us \geq \liminf_{n \to \infty}  \rho_t(-\rho_{t+1}(-H_{\theta^n})) \geq  \rho_t(-\us[t+1]),
\]
since the conditional risk measure $\rho_t$ has the Fatou property.\fintheo\\

In the next theorem we construct a stopping time of $t$-maximal risk. The proof will  involve the property of time-consistency of the  convex risk measure $\Phi$. 
%****************************************************************************************
\nt{rcace}{
%****************************************************************************************
Let $t \in \{0,1,\cdots,T\}$. Then, the stopping time defined by
\[
\tau^{\uparrow}_t:= \inf \{s \geq t \mid H_{s}= \us[s] \},
\]
is of $t$-maximal risk.  Thus,
\eq{labelequationoptimality}
{
\rho_t(-H_{\tau^{\uparrow}_{t}})=\us.
}
}
Proof. Clearly we have $R(\tau^{\uparrow}_t \leq T) =1$, since $\us[T]=H_T$, due to the normalization property \eqref{labelequationnormalization}. Note that $\tau^{\uparrow}_T=T$. By way of an induction argument, we get
\eq{labelequationinductionargument}
{
\us[t]= H_{t}\vee  \rho_t (-H_{\tau^{\uparrow}_{t+1}}),
} 
due to the recursive formula \eqref{labelequationbwcomputation} of Lemma  \ref{labellemmarecurusenvelop}.  The proof will be finished after we prove:
\eq{labelequationoptimalitytransitorium}
{
H_{t}\vee  \rho_t (-H_{\tau^{\uparrow}_{t+1}})=\rho_t(-H_{\tau^{\uparrow}_{t}}).
}
The following relationships are clear:
\eq{labelequationrelationshipstop}
{
\tau^{\uparrow}_{t}=1_{\{\tau^{\uparrow}_{t}= t \}} t  +  1_{\{\tau^{\uparrow}_{t} > t\}}\tau^{\uparrow}_{t+1}.
}
Now, let us define $A:=\rho_t (-H_{\tau^{\uparrow}_{t+1}})$. Then
\[
H_{t}\vee  \rho_t (-H_{\tau^{\uparrow}_{t+1}})= H_{t} 1_{\{H_{t} \geq A\}} +   A 1_{\{H_{t} < A\}},
\] 
due to \eqref{labelequationinductionargument}. Moreover, 
\[
\rho_t(-H_{\tau^{\uparrow}_{t}})= H_t 1_{\{\tau^{\uparrow}_{t} = t\}} +   \rho_t(-H_{\tau^{\uparrow}_{t+1}})1_{\{\tau^{\uparrow}_{t} > t\}},
\]
due to the localization property of Lemma \ref{labellemmalocalproperty} and the relationships \eqref{labelequationrelationshipstop}.
We then conclude \eqref{labelequationoptimalitytransitorium}, and at the same time \eqref{labelequationoptimality}, due to the following obvious set equalities:
\[
\left\{\tau^{\uparrow}_{t} = t\right\}=\{H_t=\us\}=\{H_t \geq \rho_t (-H_{\tau^{\uparrow}_{t+1}}) \}.\fintheo\\
\]
%******************************************************************************************************************************************************
\subsection{Dates of maximal risk for coherent risk measures} \label{labelsectionexamplecoherent}
%******************************************************************************************************************************************************
Let us assume that the robust representation of a time-consistent risk measure $\Phi=\{\rho_t\}_{t=0,1,\cdots,T}$ reduces to
\[
\rho_t(X)= \esssup_{P \in \mP} E_P[-X \mid \mf_t].
\]
Thus, $\rho$ is a coherent risk measure. The class $\mP$ consist of equivalent probability measures. The property of time consistency is equivalent to a stability property of the class $\mP$; see e.g., F\"ollmer and Penner \cite{Follmerpenner}, Corollary 4.12,  Delbaen\cite{Delbaenastablesets}, Theorem 6.2.\\

For $P \in \mP$, the Snell envelope at time $t$ of $H$ is defined  by
\[
U^P_t:=\esssup_{\theta \geq t} E_P[H_{\theta}\mid \mf_t].
\]
Hence 
\[
\us= \esssup_{P \in \mP} U^P_t.
\]

The stopping time:
\[
\tau^P_t:= \inf\{u \geq t \mid H_u \geq U^P_u\},
\]
is the minimal optimal stopping time of $H$ with respect to $P$ at time $t$:
\[
U^P_t=E_P[H_{\tau^P_t}\mid \mf_t];
\]
see e.g., \cite{Follmerschied}, Theorem 6.20.  In the next proposition we characterize the minimal stopping time of maximal-risk $\tau^{\uparrow}_t$ in terms of the family $\{\tau^P_t\}_{P \in \mP}$.

\np{labelpropositionuppenvelostimes}
{
The minimal stopping time of maximal-risk $\tau^{\uparrow}_t$ satisfies
\[
\tau^{\uparrow}_t= \esssup_{P \in \mP} \tau^P_t.
\]
}
Proof. The inequality $\geq$ holds true, since  $U^P \leq \us$ for each $P \in \mP$. Now let us define
\[
A:=\{\tau^{\uparrow}_t > \esssup_{P \in \mP} \tau^P_t\}.
\]
By way of contradiction, assume that $R(A)>0$. There exist a sequence of probability measures $P^i \in \mP$ such that
\[
U^{P^i}_{\tau^{\uparrow}_t} \nearrow \us[\tau^{\uparrow}_t],
\]
see \cite{Follmerschied}, Lemma 6.50. Let $j^0$ be such that $t \leq j^0 \leq T$ and 
\[
A \cap \{\tau^{\uparrow}_t=j^0\}
\]
has positive probability. If $j^0=t$  we easily generate a contradiction. Let us assume that $j^0>t$.  There exists an index $j^1 \in \{t, t+1, \cdots,j^0-1\}$ and a subsequence $i^k$ such that the event
\[
B:= A \cap \bigcap_{k=1}^{\infty}\{\tau^{P^{i^k}}_t = j^1\},
\]
has $R$-positive probability. In the event $B$ we get:
\[
H_{j^1}=H^{P^{i}}_{j^1} \nearrow \us[j^1],
\]
a clear contradiction with the definition of the stopping time $\tau^{\uparrow}_t$.\finpro

%*******************************************************************************************************************************************************
\subsection{Robust detection of maximal risk} \label{labelsectionrobuststatistics}
%*******************************************************************************************************************************************************
Let us consider the time-consistent risk measures $\Phi^1$ and $\Phi^2$.  Let us denote by $U^1$ ($U^2$) the upper Snell envelope of  $H$ with respect to $\Phi^1$ ($\Phi^2$).  Let us assume that the risk measures satisfy the order relation 
\eq{labelequationorderrelationriskmeasures}
{
\Phi^1 \leq \Phi^2,
}  
in the sense that 
\[
\rho^1_t(X) \leq \rho^2_t(X), \thinspace R-a.s.,
\]
for each $X \in L^1(\mf,R)$ and $t \in \{0,1, \ldots, T\}$. Then it is clear that
\eq{labelequationorderrelationshipsnellenvelopes}
{
U^1_t \leq U^2_t,  \thinspace R-a.s.
}
For $i=1,2$, the  minimal stopping time of $0$-maximal risk  with respect to $\Phi^i$ is given by 
\[
\tp^i:=\inf\{t\geq 0\mid U^i_t=H_t\}, 
\]
due to Theorem  \ref{rcace}. The relationship
\eq{labelequationorderrelationriskmeasuresstoppingtimes}
{
\tp^2 \leq \tp^1,
}
holds true due to \eqref{labelequationorderrelationshipsnellenvelopes}. It has an interesting interpretation in the following context. \\

Let $\mq$ be a class of probability measures which defines a time-consistent coherent  risk measure in $L^1$, say $\Phi^1$.  If we  interpret the class $\mq$ from the point of view of choice theory, as a family of priors, then the stopping time $\tp^1$ solves  a robust problem of monitoring. We express our knowledge about the ``exactness'' of a model $Q \in \mq$ with the penalty  $\alpha(Q)$. Assume that the pair $(\mq,\alpha)$ induces a time-consistent convex risk measure $\Phi^2$. The order relationship \eqref{labelequationorderrelationriskmeasures} holds true. Thus, dates of maximal risk for the risk measures $\Phi^1$ and $\Phi^2$ satisfy the relationship \eqref{labelequationorderrelationriskmeasuresstoppingtimes}. Loosely speaking, the penalty function $\alpha$ has the effect of a more exact and early detection of maximal risk.
%-----------------------------------------------------------------

\end{document}